\documentclass[a4paper,11pt]{article}
\usepackage{jheppub} 
\usepackage{lineno}
\usepackage{graphicx}
\usepackage{amsmath, amssymb, amsthm, fancyhdr,epsfig,slashed, mathrsfs}
\usepackage{svg}
\usepackage{subfigure}
\usepackage{natbib}
\usepackage{float}
\usepackage{lipsum}
\usepackage{mathtools} 
\usepackage{setspace}
\usepackage{tikz}
\usepackage{tikz-feynman}
\usepackage{tikzsymbols}
\tikzfeynmanset{compat=1.1.0}
\tikzfeynmanset{warn luatex=false}
\usepackage{xcolor,hyperref}
\usepackage{caption}
\usepackage{subcaption}
\usepackage{comment}
\renewcommand{\thesection}{\arabic{section}}

\DeclareRobustCommand{\orcidicon}{
	\begin{tikzpicture}
	\draw[lime, fill=lime!90!black] (0,0) 
	circle [radius=0.2] 
	node[white] {{\fontfamily{qag}\selectfont \tiny ID}};
	\draw[white, fill=white] (-0.0625,0.095) 
	circle [radius=0.007];
	\end{tikzpicture}
	\hspace{-2mm}
}
\foreach \x in {A, ..., Z}{\expandafter\xdef\csname orcid\x\endcsname{\noexpand\href{https://orcid.org/\csname orcidauthor\x\endcsname}
			{\noexpand\orcidicon}}
}

\newcommand{\AddrIITG}{Department of Physics, Indian Institute of Technology Guwahati, Assam 781039, India}
\title{Dark matter from axion and small neutrino mass}
\author{Shivam Gola\orcidA{} }
\affiliation{\AddrIITG}
\emailAdd{shivamg.sk@rnd.iitg.ac.in}
\abstract{We explore a KSVZ-like extension of the Standard Model with a Dirac fermion and three right-handed neutrinos. PQ symmetry allows the Dirac mass for neutrinos and prevents the Majorana mass. A $\mathcal{Z}_2$ symmetry guarantees the stability of Dirac fermion dark matter. The breakdown of PQ symmetry generates the QCD axion at a high scale. The fermion dark matter relic abundance arises from the UV-freeze-in mechanism through the axion portal. We determine the fermion DM relic by solving the coupled Boltzmann equations and finding the allowed parameter space using the relic density constraints. Having determined the allowed parameter space for fermion DM, we also look for non-thermal axion production schemes to seek the two DM possibility. We find that FIMP alone is a suitable dark matter that is not excluded while considering several current bounds and future sensitivities on axion and dark matter. Our study highlights the interlinking of dark matter, axion, and neutrinos while addressing the strong CP problem and small neutrino masses.}
\begin{document}
\maketitle
\flushbottom
\section{Introduction}
\label{sec:intro}
The numbers of independent astrophysical observations have confirmed the existence of dark matter (DM)~\cite{1937ApJ....86..217Z, Rubin1980RotationalPO, Clowe_2006, 1979Natur.279..381W, 1965ApJ...142..419P, Hinshaw:2012aka, Planck:2018vyg}. DM does not interact with light that makes them invisible however it plays a significant role in the large-scale structure formation of our universe. The sole observable here is the relic density bound in eq.~\ref{eq:planck} from the Planck satellite data~\cite{Planck:2018vyg}.
\begin{align}
\Omega_{\text{DM}} h^2 = 0.12 \pm 0.001.
\label{eq:planck}
\end{align}
DM abundance is nearly five times the normal matter, yet its particle composition and interactions are mostly unknown. The standard model (SM) cannot explain DM, therefore, several well-motivated beyond standard model (BSM) scenarios suggest a suitable candidate for DM \cite{Bertone:2004pz,Profumo:2019ujg,Merle:2017jfn,Jungman:1995df}. Weakly Interacting Massive Particles (WIMPs)~\cite{Profumo:2019ujg,Schumann_2019} have been a popular candidate for DM as they naturally explain the observed dark matter density through the process called freeze-out mechanism~\cite{Kolb:1990vq}. However, WIMPs are not detected in any experimental searches such as direct detection~\cite{PandaX-II:2016vec, LUX:2016ggv,Akerib:2018lyp,XENON:2018voc,Aprile:2020vtw}, indirect detection~\cite{HESS:2018cbt,2016} and collider e.g. Large Hadron Collider (LHC)~\cite{2018,CMS:2018mgb} etc. Feebly Interacting Massive Particles (FIMPs)~\cite{Hall_2010,Bernal_2017,Belanger:2018ccd} is an interesting alternative to the popular WIMP candidate. FIMP interacts with SM or dark sector (DS) particles through a very small coupling ($\lesssim {\cal O}(10^{-12} - 10^{-10})$). Consequently, FIMP never achieves thermal equilibrium with the bath particles in the early universe. However, it produces non-thermally through the decay or annihilation of the mother particles. As time progresses, the initially negligible number density of FIMP increases and eventually stabilizes due to Boltzmann suppression, leading to the correct DM abundance. This production process is called the freeze-in mechanism. The freeze-in scenario is broadly classified into two categories: 1. Infra-red (IR) freeze-in is significant at lower temperatures, and 2. Ultra-violet (UV) freeze-in occurs at higher temperatures, such as the reheating temperature of the Universe. \\
The small mass of neutrinos highlights another shortcoming of SM, as confirmed by neutrino oscillation experiments~\cite{ParticleDataGroup:2020ssz, Aghanim:2018eyx, Lattanzi:2017ubx}. This oscillation data also indicates that at least two of the three neutrinos are massive, while they are assumed to be massless in SM. To generate mass for neutrinos, one can simply add three right-handed neutrinos (RHNs) that can mix with active neutrinos through the Yukawa coupling similar to other SM fermions, resulting in the Dirac mass. \\
Now taking a slight digression, the presence of a non-vanishing CP violating $\theta$ parameter in the quantum chromodynamics (QCD) sector implies the Strong CP problem~\cite{Peccei:1996ax,Kim:2008hd,Hook:2018dlk,Lombardo:2020bvn,Irastorza_2018}. The effective $\theta$-parameter can range from 0 to 2$\pi$; however, $\mid \theta \mid \lesssim 10^{-10}$, from the measurement of neutron electric dipole moment (EDM). The dynamical solution to the strong CP problem is by Peccei–Quinn~(PQ)~\cite{Peccei:2006as, Weinberg:1977ma,Wilczek:1977pj}, which requires a pseudo-Nambu-Goldstone boson, the axion, which relaxes the $\theta$-term. Axions acquire a non-zero mass from QCD dynamics, which is inversely proportional to the axion decay constant \( f_a \). In the PQWW model, the decay constant is related to the SM Higgs vacuum expectation values (VEVs)~\cite{Peccei:2006as}, thus, it tightly constrains the solution. In invisible axion models e.g. KSVZ~\cite{Kim:1979if,Shifman:1979if}, DFSZ~\cite{Dine:1981rt} etc. the axion scale \( f_a \) is at significantly higher scale. In particular, the KSVZ model includes a complex singlet scalar and two colored quarks, all charged under a new global PQ symmetry. Spontaneous breaking of the global symmetry addresses the Strong CP problem and results in a new particle, the axion. We take inspiration from KSVZ-type models for constructing our model. \\
Several BSM models address these above-mentioned issues individually or collectively~\cite{Ma:2006km,Gola:2022nkg,Nomura_2009,Gola:2021abm,Salvio_2015,Carvajal_2017,Ballesteros_2017,Peinado_2020,ghosh2023axionlikeparticlealpportal,Bharucha_2023,delaVega:2020jcp,chao2022axionlikedarkmattertypeii,Berezhiani1991,2306.03128}. We revisit dark matter, neutrino mass, and the Strong CP problem with a minimal model which interconnects these three problems. In our model, we add a pair of quarks, a complex scalar, a Dirac fermion, and three RHNs, all are charged under the new global PQ symmetry. We also introduce a new Higgs-like scalar with a non-zero PQ charge, which enables the Yukawa coupling for neutrinos. The tree-level Lagrangian is invariant under global symmetry except for the anomaly in the QCD sector. The complex scalar spontaneously breaks the PQ symmetry, which generates mass for the heavy quarks, Dirac fermions, and Dirac neutrinos. The imaginary parts of all scalars combine, and one of the components is identified as the axion. Axion couples to gluon, photon, and neutrinos due to pseudo-scalar mixing. Lastly, the Dirac fermion is protected by an additional $\mathcal{Z}_2$ symmetry, however, it may be possible that a subgroup of PQ symmetry remains unbroken after spontaneous symmetry breaking~(SSB), which stabilizes the Dirac fermion. In either case, the Dirac fermion is a suitable candidate for DM in our model. Additionally, Dirac fermion interacts with SM through the axion portal, with interaction strength scaled by the axion decay constant \( f_a^{-1} \). Typically, $f_a > 10^8$ GeV is inferred from various searches~\cite{Raffelt:2006cw,Friedland:2012hj,Ayala:2014pea,Jaeckel_2016,Bauer:2018uxu,Hook:2019qoh}, suggesting that Dirac fermion interacts very weakly with SM, a necessary condition for UV freeze-in, which is the main focus of this work. Axions produced from the misalignment mechanism  \cite{Duffy_2009} can also serve as DM and may imply the two DM case. \\
The paper is organized as follows: Section \ref{sec:model} outlines our model, Section \ref{sec:dmanalysis} describes the methodology and analysis of dark matter, using relic density and direct detection, and limits on axion parameter space. Additionally, we studied axions and FIMPs as dark matter together, considering various existing bounds and sensitivities. In section \ref{sec:conclusion}, we present the conclusion.
%
%
\section{The Model}
\label{sec:model}
We start by formulating the Lagrangian density for the extended sector of the minimal model, which incorporates the interactions among the fields based on the charge assignments in table~\ref{tabmod}. 
\begin{table}[htbp]
    \begin{center}
    \begin{tabular}{|c|c|c|c|c|}
        \hline
        \text{} & \textbf{SU(3)} & \textbf{SU(2)} & \textbf{U(1)$_Y$} & \textbf{U(1)$_{\text{PQ}}$} \\[5pt]
        \hline
        $Q_L$ & \textbf{3} & \textbf{1} & 0 & $\frac{x_{\Phi}}{2}$ \\[5pt]
        \hline
        $Q_R$ & \textbf{3} & \textbf{1} & 0 & $-\frac{x_{\Phi}}{2}$ \\[5pt]
        \hline
        $\Phi$ & \textbf{1} & \textbf{1} & 0 & $x_{\Phi}$ \\[5pt]
        \hline
        $H_1$ & \textbf{1} & \textbf{2} & $\frac{1}{2}$ & 0 \\[5pt]
        \hline
        $H_2$ & \textbf{1} & \textbf{2} & $\frac{1}{2}$ & $x_{\phi}$ \\[5pt]
        \hline
        $\nu^k_R$ & \textbf{1} & \textbf{1} & 0 & $x_{\phi}$ \\[5pt]
        \hline
        $\psi_L$ & \textbf{1} & \textbf{1} & 0 & $\frac{x_{\phi}}{2}$ \\[5pt]
        \hline
        $\psi_R$ & \textbf{1} & \textbf{1} & 0 & $-\frac{x_{\phi}}{2}$ \\[5pt]
        \hline
    \end{tabular}
    \end{center}
\caption{Particle and symmetry content of the minimal model where $k(=1,2,3)$ represents the family index.}
\label{tabmod}
\end{table} 
The invariant Lagrangian density for the Dirac fermion DM~($\psi$), the Yukawa interactions, and the scalar sector, based on the charge assignments given in table~\ref{tabmod} are as follows,
\begin{align}
    \mathcal{L}_{\text{DM}} = & \ \bar{\psi} \gamma^\mu \partial_\mu \psi - y_\psi (\bar{\psi}_L \psi_R \Phi + \text{h.c.}) \\
    \nonumber \\ 
    \mathcal{L}_{y} = & \ -y_{u}^{ij}\overline{q_{L}^{i}}\tilde{H}_1 u^{j}_{R} - y_{d}^{ij}\overline{q_{L}^{i}} H_1 d^{j}_{R} - y_{e}^{ij}\overline{\ell_{L}^{i}} H_1 e^{j}_{R} \nonumber \\
    & - y_Q \overline{Q}_L \Phi Q_R - y_{\nu}^{ik} \overline{\ell_{L}^{i}} \tilde{H}_2 \nu^{k}_{R} + \text{h.c.} \\
    \nonumber \\
    \mathcal{L}_{s} = & \ (D^\mu H_1)^\dagger (D_\mu H_1) + (D^\mu H_2)^\dagger (D_\mu H_2) + (\partial^\mu \Phi)^\dagger (\partial_\mu \Phi) - V(H_1,H_2,\Phi) 
\label{LModel}
\end{align}
where $\tilde{H}_{1,2} = i \sigma_2 H_{1,2}^*$, and $\sigma_2$ is the Pauli matrix and the covariant derivative defined as $ D_{\mu} = \partial_{\mu} - i g_s T^a G_{\mu}^a - i g T^a W_{\mu}^a - i g_1 Y B_{\mu}^1$. The scalar potential, $V(H_1,H_2,\Phi)$\footnote{The scalar potential $V(H_1, H_2, \Phi)$ given in eq.~\ref{eq:Vpotential1} must be bounded from below~\cite{Kannike:2012pe}, which is ensured if the following conditions are satisfied: $\lambda_{H_1} > 0, \ \lambda_{H_2} > 0, \ \lambda_{\Phi} > 0, \ \lambda_{H_2}\lambda_{\Phi} - \lambda_{H_2 \Phi}^{2} > 0, \ \text{Det}(V_{\text{quartic}}) > 0 $.}, is given by:
\begin{align}
     V(H_1,H_2,\Phi) = & -\mu_{H_1}^2 H_1^\dagger H_1 - \mu_{H_2}^2 H_2^\dagger H_2 - \mu_{\Phi}^2 \Phi^\dagger \Phi \nonumber \\
    & + \lambda_{H_1} (H_1^\dagger H_1)^2 + \lambda_{H_2} (H_2^\dagger H_2)^2 + \lambda_{\Phi} (\Phi^\dagger \Phi)^2 \nonumber \\
    & + \lambda_{H_1 \Phi} (H_1^\dagger H_1)(\Phi^\dagger \Phi) + \lambda_{H_2 \Phi} (H_2^\dagger H_2)(\Phi^\dagger \Phi) \nonumber \\
    & - \lambda_{H_1 H_2}^a (H_1^\dagger H_1)(H_2^\dagger H_2) - \lambda_{H_1 H_2}^b (H_1^\dagger H_2)(H_2^\dagger H_1) \nonumber \\
    & - \kappa H_2^\dagger H_1 \Phi + \text{h.c.}
\label{eq:Vpotential1}
\end{align}
We then parameterize the scalar fields as follows:
\begin{align}
    H_1 = \frac{1}{\sqrt{2}}
        \begin{pmatrix}
        \phi_1 + i\phi_2 \\
        v_{H_1} + h + i\phi_3
        \end{pmatrix}, \quad
    H_2 = \frac{1}{\sqrt{2}}
        \begin{pmatrix}
        \phi'_1 + i \phi'_2 \\
        v_{H_2} + h' + i \phi'_3
        \end{pmatrix}, \quad
    \Phi = \frac{1}{\sqrt{2}}(v_{\Phi} + s + i \phi)
\label{eq:vevs}
\end{align}
where \( v_{H_1}, v_{H_2}, v_{\Phi} \) denote the vevs of the Higgs doublets and the complex scalar. The symmetry breaking implies mass to the heavy quarks, \( m_Q = \frac{y_Q v_{\Phi}}{\sqrt{2}} \), Dirac fermion, \( m_\psi = \frac{y_\psi v_\Phi}{\sqrt{2}} \), and, to the neutrinos, \( m_{\nu} = \frac{y_{\nu}^{ik} v_{H_2}}{\sqrt{2}} \). Additionally, in $v_{H_2} << v_{H_1} << v_{\Phi}$ limit, one linear combination from mixing of $\phi_1 \pm i\phi_2$ and $\phi'_1 \pm i \phi'_2$ is the charged Goldstone bosons that represent the longitudinal modes of the \( W^{\pm} \) bosons, while second are the charged scalar \( H^{\pm} \) for which mixing matrix $M^2_{\pm}$ is given by.
\begin{align}
    M^2_{\pm} = 
        \begin{pmatrix}
            \frac{v_{H_2}(\lambda^b_{H_1 H_2} v_{H_1} v_{H_2} + \sqrt{2} \kappa v_{\Phi})}{2 v_{H_1}} & - \frac{\lambda^b_{H_1 H_2} v_{H_1} v_{H_2} + \sqrt{2} \kappa v_{\Phi}}{2} \\
          - \frac{\lambda^b_{H_1 H_2} v_{H_1} v_{H_2} + \sqrt{2} \kappa v_{\Phi}}{2} & \frac{v_{H_1}(\lambda^b_{H_1 H_2} v_{H_1} v_{H_2} + \sqrt{2} \kappa v_{\Phi})}{2 v_{H_2}}
        \end{pmatrix}
        \approx 
       \frac{ \kappa v_{\Phi}}{\sqrt{2}} \begin{pmatrix}
            0 & -1 \\
          -1  & \frac{v_{H_1}}{v_{H_2}}
        \end{pmatrix}
\label{eq:Hpmatrix}
\end{align}
Masses of charge scalar \( H^{\pm} \) can be found by diagonalization of eq.~\ref{eq:Hpmatrix}:
\begin{align}
    m^2_{H^{\pm}}  
    \approx \frac{\sqrt{2} \kappa v_{\Phi}}{v_{H_1} v_{H_2}} v_H^2, \quad \text{where,} \ v_H = \frac{\sqrt{v_{H_1}^2 + v_{H_2}^2}}{2}
\label{eq:Hpmass}
\end{align}
Similalry the mass matrix from mixing of the real scalars \( h, h', s \) in the limit:
\begin{align}
    M^2_H = &
        \begin{pmatrix}
            2 \lambda_{H_1} v_{H_1}^2 + \frac{\kappa v_{H_2} v_{\Phi}}{\sqrt{2} v_{H_1}} & -(\lambda^a_{H_1 H_2} + \lambda^b_{H_1 H_2}) v_{H_1} v_{H_2} - \frac{\kappa v_{\Phi}}{\sqrt{2}} & \lambda_{H_1 \Phi} v_{H_1} v_{\Phi} - \frac{\kappa v_{H_2}}{\sqrt{2}} \\
            -(\lambda^a_{H_1 H_2} + \lambda^b_{H_1 H_2}) v_{H_1} v_{H_2} - \frac{\kappa v_{\Phi}}{\sqrt{2}} & 2 \lambda_{H_2} v_{H_2}^2 + \frac{\kappa v_{H_1} v_{\Phi}}{\sqrt{2} v_{H_2}} & \lambda_{H_2 \Phi} v_{H_2} v_{\Phi} - \frac{\kappa v_{H_1}}{\sqrt{2}} \\
            \lambda_{H_1 \Phi} v_{H_1} v_{\Phi} - \frac{\kappa v_{H_2}}{\sqrt{2}} & \lambda_{H_2 \Phi} v_{H_2} v_{\Phi} - \frac{\kappa v_{H_1}}{\sqrt{2}} & 2 \lambda_{\Phi} v_{\Phi}^2 + \frac{\kappa v_{H_1} v_{H_2}}{\sqrt{2} v_{\Phi}}                      
       \end{pmatrix} \nonumber \\
        & \approx
        \begin{pmatrix}
            2 \lambda_{H_1} v_{H_1}^2  &  - \frac{\kappa v_{\Phi}}{\sqrt{2}} & \lambda_{H_1 \Phi} v_{H_1} v_{\Phi}  \\
             - \frac{\kappa v_{\Phi}}{\sqrt{2}} &  \frac{\kappa v_{H_1} v_{\Phi}}{\sqrt{2} v_{H_2}} & \lambda_{H_2 \Phi} v_{H_2} v_{\Phi} - \frac{\kappa v_{H_1}}{\sqrt{2}} \\
            \lambda_{H_1 \Phi} v_{H_1} v_{\Phi}  &  \lambda_{H_2 \Phi} v_{H_2} v_{\Phi} - \frac{\kappa v_{H_1}}{\sqrt{2}} & 2 \lambda_{\Phi} v_{\Phi}^2                           
        \end{pmatrix}
\label{eq:Higgsmatrix}
\end{align}
Scalar matrix~\ref{eq:Higgsmatrix} can be diagonalized, resulting in the mass eigenstates for the real scalar fields \( h_1, h_2, h_3 \). Lastly, the neutral gauge boson mixing matrix is given by: 
\begin{align}
    M_V^2 =
        \begin{pmatrix}
            \frac{1}{4}g_1^2 v_H^2 & -\frac{1}{4} g_1 g_2 v_H^2 \\
            -\frac{1}{4} g_1 g_2 v_H^2 & \frac{1}{4} g_2^2 v_H^2
        \end{pmatrix}
\label{eq:GBmatrix}
\end{align}
This can be diagonalized to yield the mass eigenstates for the photon (\(A\)) and \(Z\)-boson, along with their respective masses:
\begin{align}
    M_A = 0, \quad M_Z^2 = \frac{g_1^2 + g_2^2}{4} v_H^2
\label{eq:mass_boson}
\end{align}
%
%
\subsection{Axion interactions}
\label{sec:axion}
The imaginary parts of the scalar fields, \( \phi_3, \phi'_3, \phi \), mix, and one component becomes the Goldstone boson of the \( Z \) boson, while the remaining two mass eigenstates are \( a' \) and \( a \). The absorbed Goldstone boson is given by,
\begin{align}
    z^0 = \frac{v_{H_1} \phi_3 + v_{H_2} \phi'_3}{2 v_H}.
\end{align}
To ensure the Goldstone boson does not mix with the axion, we require~\cite{Srednicki:1985xd, Baek_2020},
\begin{align}
    a = \frac{X_{H_1} v_{H_1} \phi_3 + X_{H_2} v_{H_2} \phi'_3 + X_{\Phi} v_{\Phi} \phi}{f_a},       
\end{align}
where $f_a = x_{\Phi} \sqrt{v_{H_2}^2 + v_{\Phi}^2}$, and the effective charges are given by,
\begin{align*}
    X_{H_1} = - x_{\Phi} \frac{v^2_{H_2}}{4 v^2_{H}}, \quad X_{H_2} = x_{\Phi} \frac{ v^2_{H_1}}{4 v^2_{H}}, \quad X_{\Phi} = x_{\Phi}.
\label{couplings1}
\end{align*}
Although, axion is massless at tree level, it can get mass through non-perturbative effects of QCD at low energy,

\begin{equation}
    m_a = \frac{m_{\pi} f_{\pi}}{f_a} \sqrt{\frac{z}{(1+z)(1+z+w)}}
\label{mass5}
\end{equation}
where \( z = m_u / m_d \) and \( w = m_u / m_s \). The axion couples to gluons and photons due to the anomaly \cite{Irastorza_2018},
\begin{align}
    -\mathcal{L}_{a-boson} = \frac{\alpha_s}{8\pi f_a} a G^{a\mu\nu} \tilde{G}^a_{\mu\nu} + \left( \frac{E}{N} - \frac{2}{3} \frac{4+z+w}{1+z+w} \right) \frac{\alpha_{em}}{8\pi f_a} a F^{\mu\nu} \tilde{F}_{\mu\nu},
\label{eq:axion-boson}    
\end{align}
where the EM-color anomaly ratio, \( \frac{E}{N} = e^2_Q \) is calculated in our model, which vanishes as the heavy quark \( Q \) is SM singlet as described in table~\ref{tabmod}. Similarly, the axion couples to neutrinos via,
\begin{align}
    \mathcal{L}_{a\nu} = X_{H_2} \frac{\partial_{\mu}a}{f_a} \left( \bar{\nu} \gamma^{\mu} \gamma^5 \nu \right).
\label{eq:axion-neutrino} 
\end{align}
Finally, the axion interacts with the heavy quark \( Q \) and fermion dark matter \( \psi \) through,
\begin{align}
    \mathcal{L}_{aQ} &= X_{\Phi} \frac{\partial_{\mu}a}{f_a} \left( \bar{Q} \gamma^{\mu} \gamma^5 Q \right), \\
    \mathcal{L}_{a\psi} &= X_{\Phi} \frac{\partial_{\mu}a}{f_a} \left( \bar{\psi} \gamma^{\mu} \gamma^5 \psi \right).
\label{eq:axion-dm} 
\end{align}
%
%
\section{Dark Matter Analysis}
\label{sec:dmanalysis}
In this section, we outline the methodology for calculating number density using the Boltzmann equation and then analyze the feasible parameter space for FIMP against various constraints. We set the vevs $v_{H_2} = 10^{-9}$ GeV, and $v_H = 246$ GeV, with very high $v_{\Phi}$, which ensure the correct masses for SM fermions and meet the requirements for FIMP production. This scaling also results in small neutrino masses, while heavy quarks and additional scalars become massive. We choose the PQ charge \( x_{\Phi} = 1 \) throughout the analysis. Before we initiate the FIMP study, it is crucial to outline a few underlying concepts and formulations in the next subsections. 
%
%
\subsection{The general Boltzmann equations}
\label{subsec:GenBEq}
We study FIMP production using the Boltzmann equation for the Friedmann-Lema\^{i}tre-Robertson-Walker~(FRW) metric. The coupled Boltzmann equations for the evolution of number densities for the Dirac fermion ($\psi$) and axion ($a$) are as follows\footnote{We have used the principle of detailed balance i.e. $\langle \sigma_{ij \rightarrow kl} v \rangle n_{i}^{\rm{eq}} n_{j}^{\rm{eq}} = \langle \sigma_{kl \rightarrow ij} v \rangle n_{k}^{\rm{eq}} n_{l}^{\rm{eq}}$ while writing equations~\ref{eq:genBEq}.},
\begin{align}
    \nonumber \frac{d n_{\psi}}{d t} + 3 H n_{\psi} = & \sum_{\rm SM} \langle \sigma_{\psi \bar{\psi} \rightarrow \rm{SM} \ \bar{\rm{SM}} }v \rangle \left( \left( n_{\psi}^{\rm{eq}}\right)^2 - n_{\psi}^{2}\right) + \langle \sigma_{a a \rightarrow \psi \bar{\psi} } v \rangle n_{a}^{2} - \langle \sigma_{\psi \bar{\psi} \rightarrow a a } v \rangle n_{\psi}^{2} , \\
    \nonumber \frac{d n_a}{d t} + 3 H n_a = & \sum_{\rm SM} \langle \Gamma_{a \rightarrow \rm{SM} \ \rm{SM}} \rangle \left( n_{a}^{\rm{eq}} - n_{a} \right) + \sum_{\rm SM} \langle \sigma_{\rm{SM} \ \bar{\rm{SM}} \rightarrow \rm{SM} \ a }v \rangle n^2_{\rm SM} \left( 1 - \frac{n_a}{n_a^{\rm eq}} \right)  \nonumber
    \\ &    + \sum_{\rm SM} \langle \sigma_{a a \rightarrow \rm{SM} \ \bar{\rm{SM}} }v \rangle \left( \left( n_{a}^{\rm{eq}}\right)^2 - n_{a}^{2}\right) - \langle \sigma_{a a \rightarrow \psi \bar{\psi} } v \rangle n_{a}^{2} + \langle \sigma_{\psi \bar{\psi} \rightarrow a a } v \rangle n_{\psi}^{2}.
\label{eq:genBEq}
\end{align}
where the SM particle distribution function is of the equilibrium distribution at the photon temperature since they were initially in thermal equilibrium with the photon bath. The thermally averaged cross-section $\langle \sigma v \rangle$ in eq.~\ref{eq:genBEq} is derived using Maxwell-Boltzmann (MB) statistics and given in eq.~\ref{eq:Thcross-section},
\begin{align}
    \langle \sigma_{1 2 \to 3 4} v \rangle = \frac{C}{2\,T K_2(m_1/T)\,K_2(m_2/T)}\int_{s_{\rm{min}}}^\infty \sigma(s) \frac{F(m_1,m_2,s)^2}{m_1^2 m_2^2\sqrt{s}}\,K_1(\sqrt{s}/T) \ d s.
\label{eq:Thcross-section} 
\end{align}
where \( C=(1)\frac{1}{2} \), with (non-) identical initial states, \( F(m_1,m_2,s) = \frac{\sqrt{(s - (m_1 + m_2)^2)(s - (m_1 - m_2)^2)}}{2} \), and \( s_{\text{min}} = \max[ (m_1 + m_2)^2, (m_3 + m_4)^2 ] \).
The thermal average decay width for axion in eq.~\ref{eq:genBEq} can be calculated as follows:
\begin{equation}
    \langle \Gamma_a \rangle = \Gamma_a \frac{K_1(m_a/T)}{K_2(m_a/T)}.
\label{eq:ThWidth}
\end{equation}
Finally, the Hubble expansion rate is given by \( H = \sqrt{\frac{8}{3} \pi G \rho} \) and the energy density of Standard Model particles is \( \rho_{\rm{SM}} = g_{*\rho,\rm{SM}}(T) \frac{\pi^2}{30} T^4 \), where $G$ is the gravitational constant and $g_{*\rho,\rm{SM}}(T)$ represents the SM effective degrees of freedom at temperature T. 
%
%
\subsection{Freeze-in regime}
\label{subsec:FI}
\begin{figure}[htbp]
\hspace{-1cm}
    \includegraphics[width=0.52\textwidth]{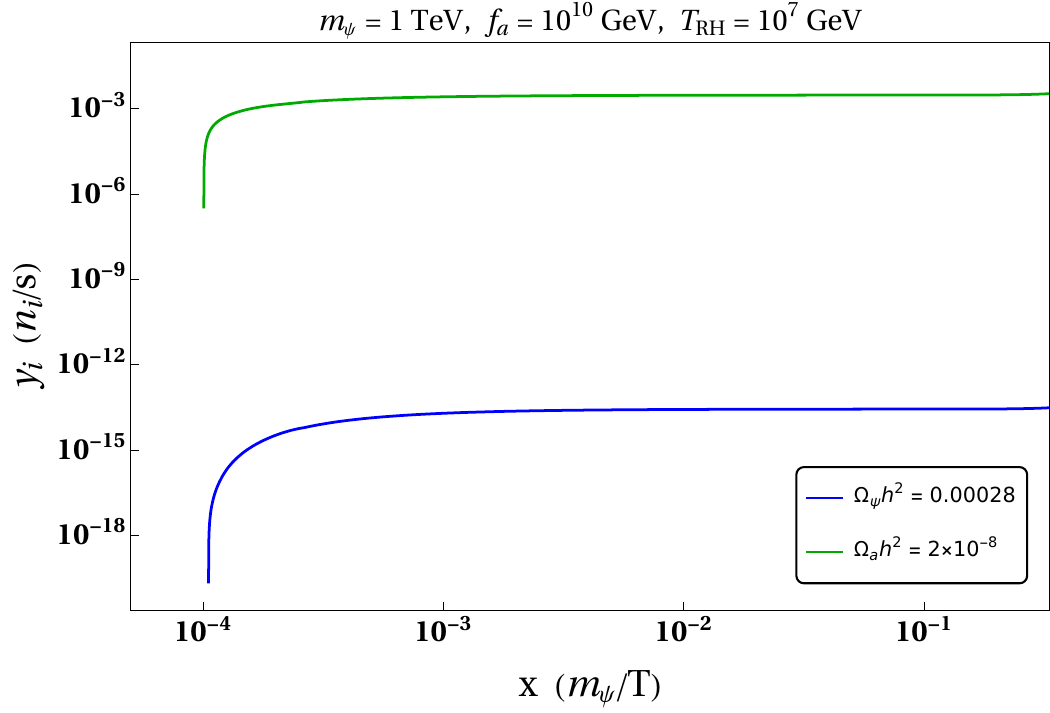} \quad 
    \includegraphics[width=0.52\textwidth]{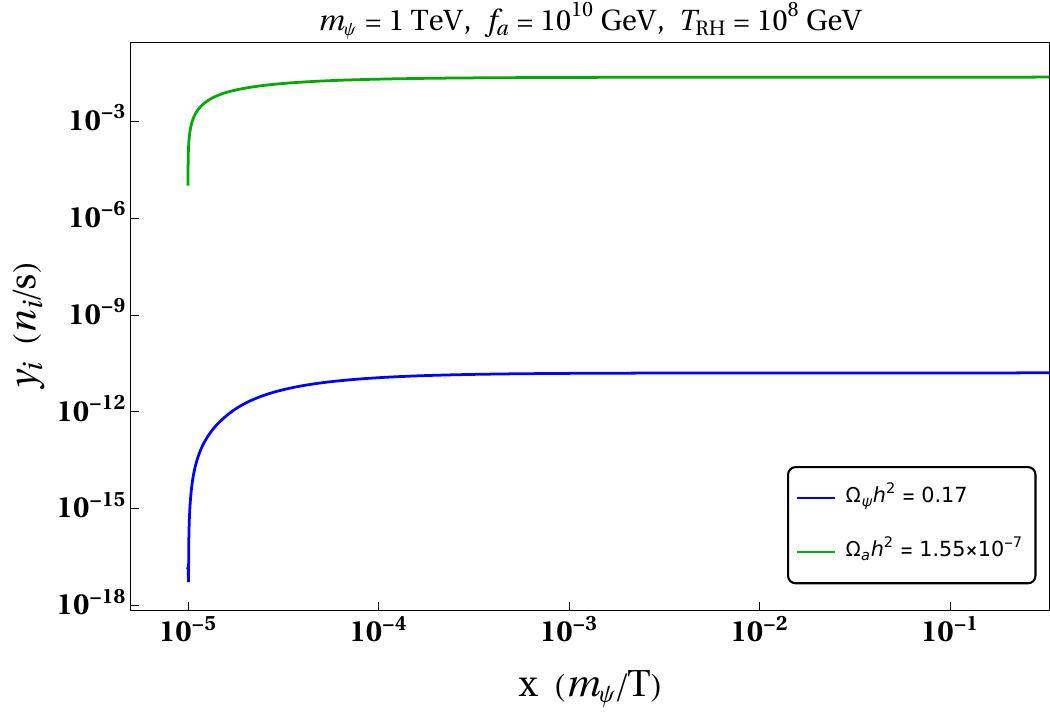}
    \caption{The figure shows how the yields (\( y_{\psi}, y_{a} \)) evolves with temperature $x (m_{\psi}/T)$ represented by blue and green curves. We set \(m_{\psi} = 1\) TeV, \(f_a = 10^{10} \) GeV and \(T_{RH} = 10^{7,8}\) GeV in left and right panels respectively.}  
\label{fig:yield}
\end{figure}

In the freeze-in regime, DM does not reach thermal equilibrium with the visible sector due to tiny couplings with SM particles. The initial small abundance of DM increases over time and freezes in when the temperature falls below the DM mass. To solve eq.~\ref{eq:genBEq}, we substitute $Y = \frac{n}{s}$ and $x = \frac{m_{\psi}}{T}$, and apply the entropy conservation $\frac{d(s a^3)}{d t} = 0$, to derive the following equations:
\begin{align}
    \nonumber s H x \frac{d y_{\psi}}{d x} = & \sum_{\rm SM} \langle \sigma_{\psi \bar{\psi} \rightarrow \rm{SM} \ \bar{\rm{SM}} }v \rangle \left( \left( n_{\psi}^{\rm{eq}}\right)^2 - s^2 y_{\psi}^{2}\right) + \langle \sigma_{a a \rightarrow \psi \bar{\psi} } v \rangle s^2 y_{a}^{2} - \langle \sigma_{\psi \bar{\psi} \rightarrow a a } v \rangle s^2 y_{\psi}^{2}
\\
    s H x \frac{d y_{a}}{d x}  = & \sum_{\rm SM} \langle \Gamma_{a \rightarrow \rm{SM} \ \rm{SM}} \rangle \left( n_{a}^{\rm{eq}} - s y_{a} \right) + \sum_{\rm SM} \langle \sigma_{\rm{SM} \ \bar{\rm{SM}} \rightarrow \rm{SM} \ a }v \rangle n^2_{\rm SM} \left( 1 - \frac{y_a}{y_a^{\rm eq}} \right)  \nonumber
    \\  &  + \sum_{\rm SM} \langle \sigma_{a a \rightarrow \rm{SM} \ \bar{\rm{SM}} }v \rangle \left( \left( n_{a}^{\rm{eq}}\right)^2 - s^2 y_{a}^{2}\right) - \langle \sigma_{a a \rightarrow \psi \bar{\psi} } v \rangle s^2 y_{a}^{2} + \langle \sigma_{\psi \bar{\psi} \rightarrow a a } v \rangle s^2 y_{\psi}^{2}.
\label{eq:genBEq1}
\end{align}
\begin{table}[htbp]
    \begin{center}
    \setlength{\tabcolsep}{10pt} 
    \begin{tabular}{|c|c|c|}
        \hline
        \rotatebox{90}{\textbf{DM - SM}}
        &              

        \begin{tikzpicture}
            \begin{feynman}
            \vertex (a);
            \vertex [above left=of a] (b) {$V$};
            \vertex [below left=of a] (c) {$V$};
            \vertex [right =of a] (d);
            \vertex [above right=of d] (e) {$\psi$};
            \vertex [below right=of d] (f) {$\bar{\psi}$};
            \diagram{
            (b) -- [boson] (a) -- [boson] (c);
            (a) -- [scalar, edge label=a] (d);
            (f) -- [fermion] (d) -- [fermion] (e);   };
            \end{feynman}
            \end{tikzpicture} 
                
        &
        
        $  g_{aVV}^2 g_{a\psi\psi}^2  $   \\
        \hline
        \rotatebox{90}{\textbf{DM - Axion}}

        & 

        \begin{tikzpicture}
            \begin{feynman}[small]
            \vertex (a);
            \vertex [above left=of a] (b) {$a$};
            \vertex [above right=of a] (c) {$\psi$};
            \vertex [below =of a] (d);
            \vertex [below left=of d] (e) {$a$};
            \vertex [below right=of d] (f) {$\bar{\psi}$};
            \diagram{
            (b) -- [scalar] (a) -- [fermion] (c);
            (a) -- [plain ] (d);
            (f) -- [fermion] (d) -- [scalar] (e);  };
            \end{feynman}
        \end{tikzpicture}
        &

        $ g_{a\psi\psi}^4 $   \\
            
        \hline
                
        &
       \begin{tikzpicture}
            \begin{feynman}[small]
            \vertex (a);
            \vertex [above left=of a] (b) {$g$};
            \vertex [above right=of a] (c) {$g$};
            \vertex [below =of a] (d);
            \vertex [below left=of d] (e) {$g$};
            \vertex [below right=of d] (f) {$a$};
            \diagram{
            (b) -- [gluon] (a) -- [gluon] (c);
            (a) -- [gluon ] (d);
            (f) -- [scalar] (d) -- [gluon] (e);  };
            \end{feynman}
            \end{tikzpicture} 
       \begin{tikzpicture}
            \begin{feynman}[small]
            \vertex (a);
            \vertex [above left=of a] (b) {$g$};
            \vertex [above right=of a] (c) {$g$};
            \vertex [below =of a] (d);
            \vertex [below left=of d] (e) {$g$};
            \vertex [below right=of d] (f) {$a$};
            \diagram{
            (e) -- [gluon] (a) -- [gluon] (c);
            (a) -- [gluon ] (d);
            (f) -- [scalar] (d) -- [gluon] (b);  };
            \end{feynman}
            \end{tikzpicture} 
        \begin{tikzpicture}
            \begin{feynman}[small]
            \vertex (a);
            \vertex [above left=of a] (b) {$g$};
            \vertex [below left=of a] (c) {$g$};
            \vertex [right =of a] (d);
            \vertex [above right=of d] (e) {$g$};
            \vertex [below right=of d] (f) {$a$};
            \diagram{
            (b) -- [gluon] (a) -- [gluon] (c);
            (a) -- [gluon, edge label=] (d);
            (f) -- [scalar] (d) -- [gluon] (e);   };
            \end{feynman}
            \end{tikzpicture} 
        \begin{tikzpicture}
            \begin{feynman}
            \vertex (a);
            \vertex [above left=of a] (b) {$g$};
            \vertex [below left=of a] (c) {$g$};
            \vertex [above right=of a] (e) {$g$};
            \vertex [below right=of a] (f) {$a$};
            \diagram{
            (b) -- [gluon] (a) -- [gluon] (c);
            (f) -- [scalar] (a) -- [gluon] (e);   };
            \end{feynman}
            \end{tikzpicture}             
        &   \\

        \rotatebox{90}{\textbf{SM - Axion}} 
                    
        &
          \begin{tikzpicture}
            \begin{feynman}[small]
            \vertex (a);
            \vertex [above left=of a] (b) {$q$};
            \vertex [below left=of a] (c) {$\bar{q}$};
            \vertex [right =of a] (d);
            \vertex [above right=of d] (e) {$g$};
            \vertex [below right=of d] (f) {$a$};
            \diagram{
            (b) -- [fermion] (a) -- [fermion] (c);
            (a) -- [gluon, edge label=] (d);
            (f) -- [scalar] (d) -- [gluon] (e);   };
            \end{feynman}
            \end{tikzpicture} 
       \hspace{1cm}
       \begin{tikzpicture}
            \begin{feynman}[small]
            \vertex (a);
            \vertex [above left=of a] (b) {$a$};
            \vertex [above right=of a] (c) {$V$};
            \vertex [below =of a] (d);
            \vertex [below left=of d] (e) {$a$};
            \vertex [below right=of d] (f) {$\bar{V}$};
            \diagram{
            (b) -- [scalar] (a) -- [boson] (c);
            (a) -- [boson ] (d);
            (f) -- [boson] (d) -- [scalar] (e);  };
            \end{feynman}
            \end{tikzpicture} 

            &    
            
            $ g^2_s g^2_{ag}, g_{aVV}^4 $    \\
            
        \hline
    \end{tabular}
    \end{center}
\caption{The relevant Feynman diagrams for axion and Dirac fermion dark matter with coupling order are shown. Here $V = \gamma, ~g$ is the photon and gluon, and $g_s$ is the strong coupling.}
\label{tab:feyn_diagrams}
\end{table}
where, $s = \frac{2 \pi^2}{45} g_{*s}(T) T^3$, is the entropy density of the Universe and $g_{*s}(T)$ is the effective degrees of freedom at temperature T. The thermal axion width and cross-section expressions are required to solve the coupled Boltzmann equations~\ref{eq:genBEq1}. These cross-sections fall into three categories: DM - SM, DM - Axion, and SM - Axion, as outlined in table~\ref{tab:feyn_diagrams}. In the table, we present the Feynman diagrams for the relevant \( 2 \to 2 \) processes: \( g g, \gamma \gamma, a a \rightarrow \psi \psi \). The expressions for the axion decay width and the annihilation cross-sections for these channels are provided in Appendices~\ref{app:axion decay widths} and~\ref{app:Thcross-section}, respectively. We used interaction rate estimates from studies in Refs.\cite{1310.6982, Graf_2011, BOLZ2001518} for the axion production processes such as $gg, q\bar{q} \rightarrow ga, \ q g \rightarrow q a$ represented as ``${\rm SM \ SM \rightarrow SM \ a}$" in equation~\ref{eq:genBEq1}. Lastly the cross section for channels $\nu \bar{\nu} \to \psi \bar{\psi}$ are suppressed by $m^2_{\nu}/f^4_a$ 
therefore these processes are not considered in the analysis. Additionally, all interactions mediated by heavy quarks Q are suppressed too and thus neglected.
\begin{figure}[htbp]
    \centering    
    \includegraphics[width=\textwidth]{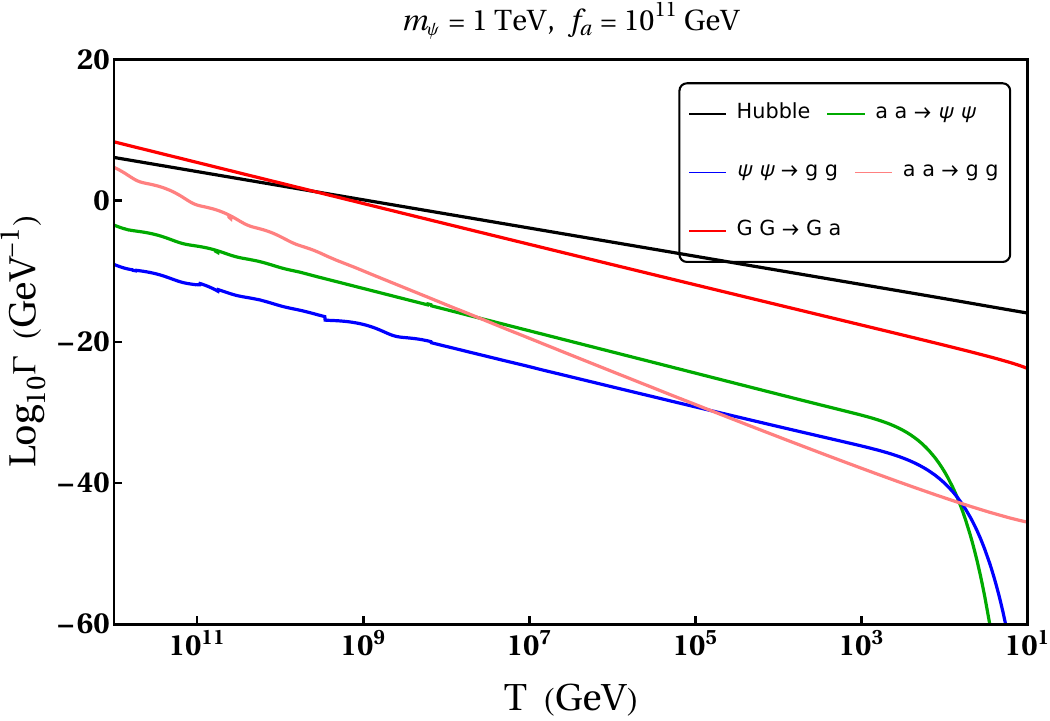}
    \caption{Hubble expansion rate and the interaction rates ($\Gamma$) for various channels with temperature are plotted for $m_{\psi}=1$ TeV, and, $f_a = 10^{11}$ GeV. Temperature dependencies of Hubble and interaction rates: DM - axion, DM - gluon, and axion - gluon are depicted in black, green, blue, pink, and red, respectively. }
\label{fig:Thrates1}
\end{figure}
The freeze-in regime occurs when the DM does not thermalize with the visible sector, i.e., the interaction rates ($\Gamma$) must drop below the Hubble expansion rate in the early Universe. The interaction rate for the process of type \( X X \rightarrow Y Y \) is as follows:
\begin{equation}
    \nonumber  \Gamma_{X X \rightarrow Y Y} = n_{X}^{\rm{eq}} \langle \sigma_{X X \rightarrow Y Y} \rangle.
\label{eq:intrates}
\end{equation}
To evaluate these rates, we choose $m_{\psi} = 1$ TeV, and, $f_a = 10^{11}$ GeV in addition to parameters fixed already in~\ref{sec:dmanalysis}. The remaining parameters can be inferred from the equations and their relations provided in sec.~\ref{sec:model}. In fig.~\ref{fig:Thrates1}, we display the interaction rates for the channels \(  \psi \psi \rightarrow g g,  a a \rightarrow g g, a a \rightarrow \psi \psi, \ \text{and,} \ G G \rightarrow G a \), alongside the Hubble expansion rate, where $G= q,g$. Interaction rates for DM-gluon, axion-gluon(annihilation), and DM-axion channels fall below the Hubble rate at high temperatures. However, axion production from quark-gluon plasma maintains axion in thermal equilibrium at temperatures above $10^9$ GeV. When the reheating temperature($T_{\rm RH}$) is lower than the axion decoupling temperature, axions remain out of thermal equilibrium, preventing dark matter from reaching thermal equilibrium as well. It's worth noting that $m_{\psi} \sim 1$ TeV suggests a Yukawa coupling of $y_{\psi} = 10^{-6}$, thus implying a small FIMP-SM interaction mediated by the higgs, hence keeping FIMP out of equilibrium. We solves the coupled boltzmann equation in eq.~\ref{eq:genBEq1} numerically for $m_{\psi} = 1$ TeV, $f_a = 10^{10}$ GeV, and, $T_{RH} = 10^{7,8}$ GeV, assuming initial FIMP and axion abundances are zero. FIMP relic is then calculated by $\Omega_\psi h^2 = m_\psi y_\psi s_0 h^2/\rho_c$, whereas, axion is very light and decouple early therefore its should be treated as thermal relic and its abundances are computed via $\Omega_a h^2 \approx \sqrt{\langle p_{a,0} \rangle^2 + m_a^2} y_a s_0 h^2/\rho_c$ \cite{Graf_2011}, where present average momentum $\langle p_{a,0} \rangle = 2.701 T_{a,0}$ and present axion temperature $T_{a,0} = 0.332 T_{0}$, where $T_0$ is present cosmic microwave background (CMB) temperature. In figure \ref{fig:yield}, we display the co-moving abundances (\( y_{\psi}, ~y_{a} \)) variation with temperature $x (m_{\psi}/T)$ for \(T_{RH} = 10^{7,8}\) GeV as in left and right panels respectively. A higher reheating temperature increases the FIMP yield \( y_{\psi} \), while the axion yield \( y_{a} \) remains relatively unchanged. This is expected, as we kept $f_a$ fixed, and it is below the $T_{\rm RH}$. We also found the axion yield is significantly contributed by the axion production channels~${\rm G \ G \rightarrow G \ a}$, which then contribute to FIMP production. Additionally, a higher $f_a$ leads to a smaller yield for both.
%
%
%
\subsection{Relic density}
\label{subsec:relic}
\begin{figure}[htbp]
\centering
    \includegraphics[width=\textwidth]{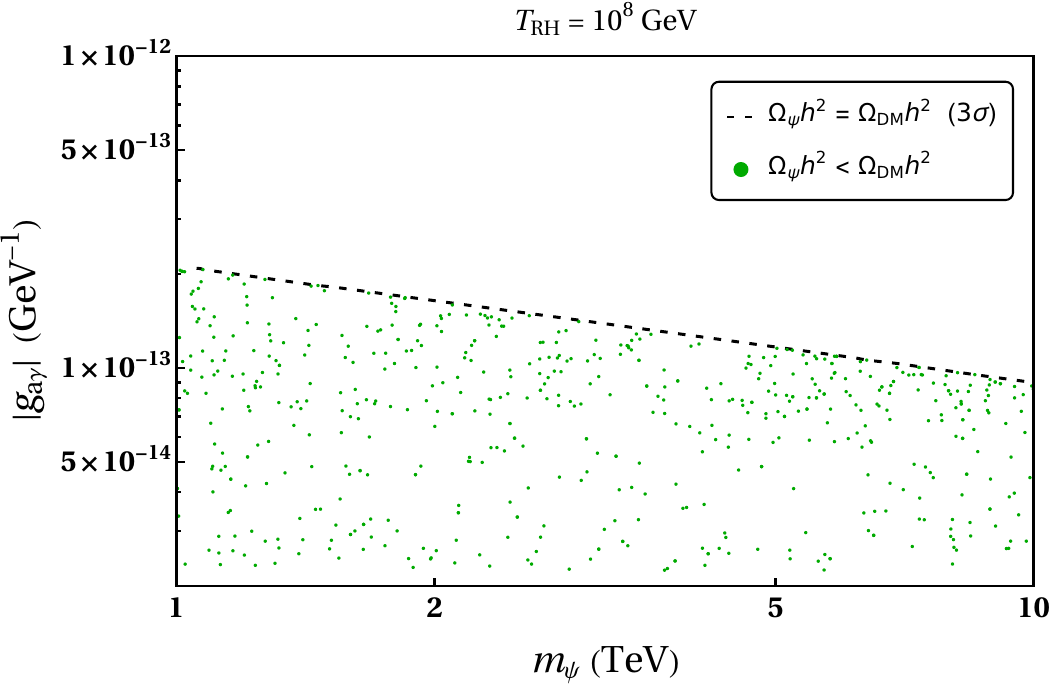}
    \caption{The panels shows the allowed region using relic constraints~\cite{Planck:2018vyg} on DM mass and axion-photon coupling ($|g_{a\gamma}|/f_a$) plane.}
\label{fig:relic_fi}
\end{figure}

\begin{figure}[htbp]
\centering
    \includegraphics[width=\textwidth]{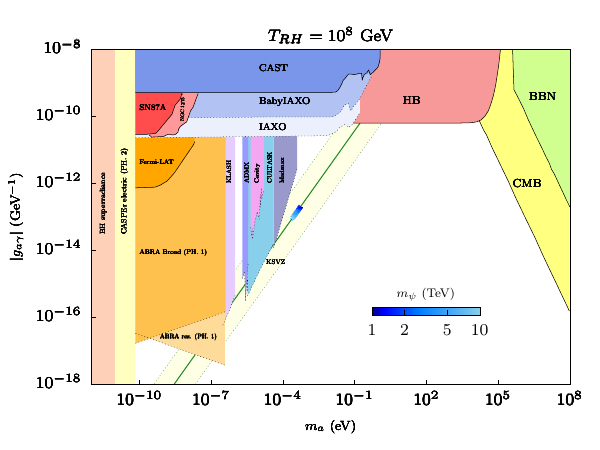}
    \caption{A summary of current bounds and future sensitivities from various experimental searches are shown on the axion mass ($m_a$) and the axion-photon coupling ($|g_{a\gamma}|$) plane. The plot displays the colored contour as the permitted parameter space for FIMP from the 3$\sigma$ range of the relic bound. }  
\label{fig:ma_cagafa1}
\end{figure}

\begin{figure}[htbp]
\hspace{-1cm}
    \includegraphics[width=0.52\textwidth]{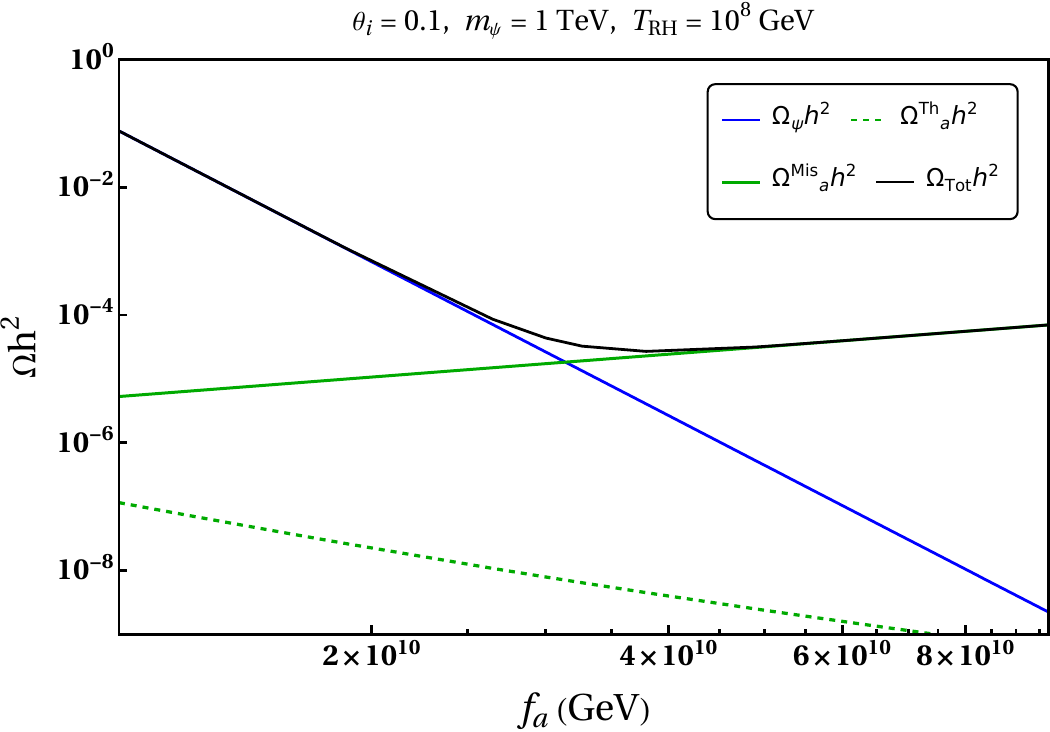}  \quad
    \includegraphics[width=0.52\textwidth]{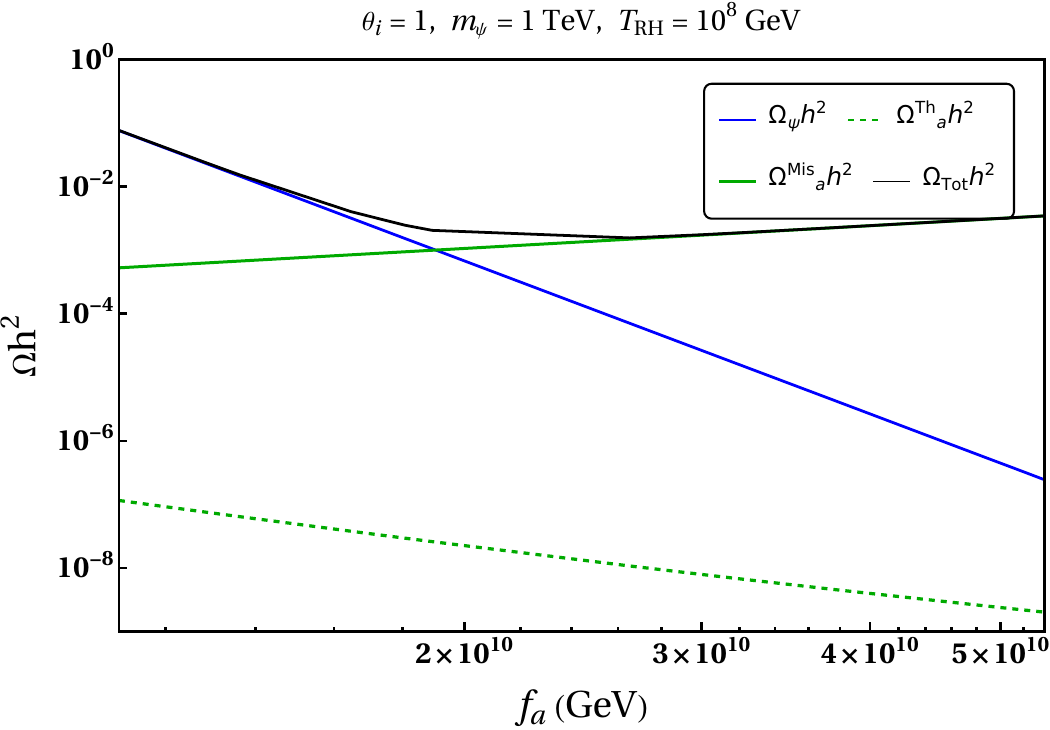}
    \caption{The plot displays the colored contour for FIMP and axion relic from thermal and non-thermal production schemes for $\theta_i = \{0.1, 1 \}$ in left and right panels, respectively. We fix $m_\psi = 1$ TeV and $T_{\rm RH} = 10^8$ GeV in both panels.  } 
\label{fig:omegas}
\end{figure}

In this section, we determine the feasible parameter space from the relic density constraint~\ref{eq:planck} on the Dirac fermion~($\psi$). We take $T_{RH} = 10^8$ GeV and $f_a > 10^{10}$ GeV, which ensures both the FIMP and axion remain out of equilibrium for $m_{\psi} \sim$ few TeV as illustrated in fig.~\ref{fig:Thrates1}. In fig.\ref{fig:relic_fi}, we display the allowed parameter space with colored data points on DM mass ($m_{\psi}$) with axion - photon coupling strength ($|g_{a\gamma}|$) plane, where, $g_{a\gamma} = \left( \frac{E}{N} - \frac{2}{3} \frac{4+z+w}{1+z+w} \right) \frac{\alpha_{em}}{2\pi f_a}$. The black dashed line in the graphs illustrates the 3$\sigma$ range from the relic bound, whereas the dark green points show the region for the underabundance of DM. We notice that a higher $m_\psi$ requires a higher $f_a$ and vice versa. Therefore, for $m_{\psi} < 1$ TeV, we require $f_a < 10^{10}$ GeV, which necessitates a check for equilibrium. Additionally, a smaller $f_a$ is subjected to experimental constraints displayed in fig.~\ref{fig:ma_cagafa1}. However, for $m_{\psi} > 10$ TeV, a higher $f_a$ is required, thus the out-of-equilibrium condition is ensured, and the allowed parameter space should follow the trend as in fig.\ref{fig:relic_fi}.
We also seek the FIMP signatures on the axion mass (\(m_a\)) and \( |g_{a\gamma}| \) plane. Figure~\ref{fig:ma_cagafa1} illustrates several bounds from astrophysical, cosmological, and other experimental searches. The solid lines represent the current experimental limits on the axion-photon coupling from CAST \cite{CAST:2017uph}, SN87A \cite{Raffelt:1987yt,Burrows:1988ah}, NGC 1275 \cite{Ajello_2016}, ADMX \cite{Braine_2020}, HB \cite{Li:2023vpv}, BBN \cite{Depta:2020wmr}, CMB \cite{Capozzi:2023xie}), etc., while the dashed lines indicate the projected sensitivities of future experiments such as CASPEr \cite{Budker_2014}, ABRACADABRA \cite{Salemi:2019xgl}, Fermi-LAT \cite{Meyer_2017}, KLASH \cite{Alesini_2023}, CULTASK \cite{Lee:2020cfj}, MADMAX \cite{Caldwell_2017}, IAXO \cite{Armengaud_2019}, BabyIAXO \cite{Ahyoune_2023}, BH superradiance \cite{Cardoso:2018tly} etc. The light yellowish band in the middle represents various QCD axion models, while the forest-green line corresponds to axion dark matter in the KSVZ model. The bluish color broad line represents the contour for the FIMP mass ranges $m_{\psi}$, extending from 1 to 10 TeV. Additionally, this bluish line falls within the 3$\sigma$ range of the relic density bound.
Now, we estimate the non-thermal production of axions, which depends on the breaking of the PQ symmetry scale and the occurrence of inflation. If PQ symmetry breaks before or during inflation, i.e., $f_a > T_{RH}$, it effectively dilutes the contributions from strings and domain wall, leaving only the misalignment contribution \cite{Chun_2014,Choi_2014}, which is as follows:
\begin{equation}  
    \Omega_a h^2 \approx \Omega_{\text{DM}} h^2 \left[ \theta_i^2 + \left( \frac{H_I}{2\pi f_I} \right)^2 \right] \left( \frac{f_a}{10^{12} \, \text{GeV}} \right)^{1.19} \left( \frac{\Lambda_{\text{QCD}}}{400 \, \text{MeV}} \right).
\label{eq:omegaax1}
\end{equation}  
Here, \(\theta_i\) is the uniform initial misalignment angle from a small patch that expanded during inflation. The parameters  \(f_I\) and \(H_I\) are the axion decay constant and Hubble parameter during inflation. The uniform axion field acquires quantum fluctuations during inflation, $\delta a \approx H_I/2\pi $. These fluctuations are subject to constraint from the isocurvature power spectrum $\mathcal{P}_a$ of axion CDM relative to the scalar power spectrum $\mathcal{P}_r$ \cite{Planck:2013jfk}.
\begin{equation}
    \frac{\mathcal{P}_a}{\mathcal{P}_r} = 4 \left(\frac{\Omega_a}{0.12}\right)^2 \frac{(H_I/2\pi)^2}{(f_I \theta_I)^2 + (H_I/2\pi)^2} \leq 0.04
\label{eq:isocurv}    
\end{equation}
Therefore, a larger $f_I~( > f_a)$ is important to suppress these fluctuations. We carefully choose $\theta_i = \{0.1, 1\}$, $H_I = 10^{14}$ GeV, $f_I = 10^{17}$ GeV, which respect the isocuravture bound \cite{Planck:2013jfk} using equation~\ref{eq:isocurv}. It is straightforward to calculate the axion relic density using eq.~\ref{eq:omegaax1} as $\Omega_a^{\rm Mis} h^2$, while the thermal axion relic $\Omega_a^{\rm Th} h^2$ is calculated using eq.~\ref{eq:genBEq1}. The total relic abundance: $\Omega_{\rm Tot} h^2$ is simply the scalar sum of FIMP and axion relics. $\Omega_a^{\rm Mis} h^2$ depends on the initial misalignment angle $\theta_i$, which implies the non-thermal axion production can be significant as FIMP, as shown in fig.~\ref{fig:omegas} while respecting the isocurvature bounds. Finally, if PQ symmetry breaks after inflation ($f_a < T_{RH}$), axions can be produced through the misalignment mechanism, strings, and domain walls \cite{Duffy_2009,kawasaki2015axion,hiramatsu2012production,Ringwald_2016}. However, in this case, axion from interaction channels can thermalize due to a smaller $f_a$, and that may cause problems for our FIMP analysis, so we exclude this case from our analysis.
%
%
Lastly, the stringent direct detection constraints from experiments such as, LUX~\cite{LUX:2016ggv} and XENON1T~\cite{XENON:2018voc} can be bypassed due to a smaller scattering cross-section resulting from the axion mediation~\cite{Cheng:2012qr,Dolan:2014ska,Gola:2021abm}. The DM-nucleon scattering cross-section has a $q^4$ momentum suppression due to $\gamma_5$ in the scattering matrix. Additionally, the scattering cross-section is further suppressed due to $f_a^{-2}$ factor.

%
%
\section{Conclusion}
\label{sec:conclusion}
We study a fermionic DM model with axion as the mediator in a KSVZ-like extension of SM. In this model, we conduct a detailed analysis of the interplay between DM, axion, and neutrino mass generation using the Peccei-Quinn (PQ) symmetry. The introduction of axions dynamically resolves the strong CP problem, while small neutrino masses are generated due to the PQ charged Higgs-like doublet. We emphasize the limitations of WIMP and present FIMPs as a compelling alternative. The high-scale physics of KSVZ-like axion and its coupling with fermion DM suggests the UV freeze-in mechanism for its production, which also evades the stringent direct detection bounds. We examine axion and FIMP as DM separately and together while considering several existing bounds and projected experimental limits on axion mass and its coupling with the photon. This simple extension to SM can provide good candidates to DM, generate Dirac mass to neutrinos, and solve the Strong CP problem; by interlinking them, it may be a promising extension to KSVZ-type models.
%
%
\acknowledgments
This work is supported by the Science and Engineering Research Board (SERB), Government of India grant CRG/2022/000603. We thank Dr. Debasish Borah for his insightful discussions and guidance, which were essential to completing this work. We thank the referee for the critical comments, which helped us to improve the work. We also thank IMSc, Chennai, for their support during my postdoctoral position search, which facilitated the beginning of this work.

%
%
\appendix
%
%
\section{Axion Decay Widths}
\label{app:axion decay widths}
The relevant axion decay width expressions are as follows:
\begin{align*}
\label{app:Thaxwidth}
\Gamma_{a \rightarrow g g}= \frac{4 m_a^3 g_{ag}^2 }{\pi },\,\ 
\Gamma_{a \rightarrow \gamma \gamma}=\frac{m^3_a g_{a\gamma}^2 }{32\pi },\,\ 
\Gamma_{a \rightarrow \psi \bar{\psi}}=\frac{X_{\Phi}^2 m_a m_{\psi}^2 \sqrt{1-\frac{4m^2_{\psi}}{m_a^2}} }{8\pi f_a^2} 
\end{align*}
where $g_{ag} = \frac{\alpha_s}{8\pi f_a} $ and $g_{a\gamma} = \left( \frac{E}{N} - \frac{2}{3} \frac{4+z+w}{1+z+w} \right) \frac{\alpha_{em}}{2\pi f_a}$ with $\alpha_s$ and, $\alpha_{\text{em}}$ as  strong and electromagnet coupling.
%
%
\section{Annihilation cross sections}
\label{app:Thcross-section}
The relevant cross-section expressions for many annihilation channels are as follows:
\begin{align*}
& \sigma_{gg \rightarrow \psi \psi} = \frac{X^2_{\Phi} g_{ag}^2 m^2_{\psi} s^2 \sqrt{1-\frac{4m^2_{\psi}}{s}} }{8 \pi f^2_a(m^2_{a}-s)^2 }, \quad \quad 
\sigma_{\gamma \gamma \rightarrow \psi \psi} = \frac{ X^2_{\Phi} g_{a\gamma}^2 m^2_{\psi}s^2 \sqrt{1-\frac{4m^2_{\psi}}{s}} }{16 \pi f^2_a(m^2_{a}-s)^2 } \\ 
& \sigma_{a a \rightarrow \psi \psi} = \frac{2 X_{\Phi}^4 m_{\psi}^2}{\pi f_a^4 s (s - 4 m_a^2)} \Bigg\{ \frac{\sqrt{(s - 4 m_a^2) (s - 4 m_{\psi}^2)} (-4 m_a^2 m_{\psi}^2 s + m_{\psi}^2 s^2 + m_a^4 (s - 2 m_{\psi}^2))}{m_a^4 - 4 m_a^2 m_{\psi}^2 + m_{\psi}^2 s} + \\ & \quad \frac{2 m_{\psi}^2 (2 m_a^4 - 4 m_a^2 s + s^2) \log\left(\frac{s - 2 m_a^2 - \sqrt{(s - 4 m_a^2) (s - 4 m_{\psi}^2)}}{s - 2 m_a^2 + \sqrt{(s - 4 m_a^2) (s - 4 m_{\psi}^2)}}\right)}{s - 2 m_a^2} \Bigg\} \\
& \sigma_{a a \rightarrow g g} = \frac{8 g_{ag}^4}{\pi s (s - 2 m_a^2) (s - 4 m_a^2)} \Bigg\{ \sqrt{s (s - 4 m_a^2)} (-12 m_a^6 + 14 m_a^4 s - 14 m_a^2 s^2 + 5 s^3) + \\ & \quad  4 m_a^4 (3 m_a^4 - 4 m_a^2 s + s^2) \log\left(\frac{s - 2 m_a^2 - \sqrt{s (s - 4 m_a^2)}}{s - 2 m_a^2 + \sqrt{s (s - 4 m_a^2)}}\right) \Bigg\} \\
& \sigma_{a a \rightarrow \gamma \gamma} = \frac{g_{a\gamma}^4}{256 \pi s (s - 2 m_a^2) (s - 4 m_a^2)} \Bigg\{ \sqrt{s (s - 4 m_a^2)} (-12 m_a^6 + 14 m_a^4 s - 14 m_a^2 s^2 + 5 s^3) + \\ & \quad 4 m_a^4 (3 m_a^4 - 4 m_a^2 s + s^2) \log\left(\frac{s - 2 m_a^2 - \sqrt{s (s - 4 m_a^2)}}{s - 2 m_a^2 + \sqrt{s (s - 4 m_a^2)}}\right) \Bigg\} 
\end{align*}
\bibliographystyle{JHEP}
\bibliography{bibitem} 
\end{document}